\documentclass[a4paper]{jpconf}

\usepackage{epsfig}
\usepackage{graphicx,latexsym}
\usepackage{dcolumn}
\usepackage{amssymb,amsmath,bm}

\usepackage[dvips]{color}

\usepackage{hyperref}
\hypersetup{
    pdfnewwindow=true,       
    colorlinks=true,         
    linkcolor=blue,          
    citecolor=blue,          
    filecolor=magenta,       
    urlcolor=black           
}

\def\eq#1{Eq.\ (\ref{#1})}
\def\mb#1{\mbox{\boldmath$#1$}}
\def\fig#1{figure\ \ref{#1}}

\begin{document}
\title{
Finger-gate manipulated quantum transport in Dirac materials\\
}
\author{Ioannis Kleftogiannis}
\ead{ph04917@nctu.edu.tw}
\address{Department of Electrophysics, National Chiao Tung
University,  Hsinchu 30010, Taiwan, Republic of China}
\author{Chi-Shung Tang}
\ead{cstang@nuu.edu.tw}
\address{Department of Mechanical Engineering, National United University,
 Miaoli 36003, Taiwan, Republic of China}
\author{Shun-Jen Cheng}
\address{Department of Electrophysics, National Chiao Tung
University,  Hsinchu 30010, Taiwan, Republic of China}

\begin{abstract}
We investigate the quantum transport properties of
multichannel nanoribbons made of materials described by the Dirac
equation, under an in-plane magnetic field. In the low energy
regime, positive and negative finger-gate potentials allow the
electrons to make intra-subband transitions via hole-like or
electron-like quasibound states (QBS), respectively, resulting in
dips in the conductance. In the high energy regime, double dip
structures in the conductance are found, attributed to spin-flip or
spin-nonflip inter-subband transitions through the QBSs. Inverting
the finger-gate polarity offers the possibility to manipulate the
spin polarized electronic transport to achieve a controlled
spin-switch.

\end{abstract}

\section{Introduction}

Dirac materials described by the Dirac equation for relativistic
particles is a quickly developing field in low-dimensional
mesoscopic systems that provides a wide field for both fundamental
theoretical research and applications. One of the most well known
examples is graphene where the electrons at the Fermi energy behave
as relativistic massless particles \cite{novoselov2004,Neto}.
Graphene is a promising material for integration in nanoscale
devices due to its high carrier mobility, and therefore has been
considered for implementation in semimetals \cite{novoselov2004},
nanoelectronics\cite{Berger2004}, coherent devices
\cite{Berger2006}, and field-effect transistors \cite{Wu2008}.
However, its application on nanoelectronics is limited due to the
lack of band gap at the Fermi energy. A well-established method to
overcome this problem is by patterning graphene sheets into long
stripes known as nanoribbons with varying widths using planar
technologies of electron beam lithography and etching
\cite{Son-PRL2006,Fagas2004,Peres2006,Wang2008,Lin2008}. For example
armchair graphene nanoribbons possess a gapped energy subband
structure tunable by controlling the nanoribbon width
\cite{nakada,tao}. For some other celebrated examples,
a considerable gap can be created by embending graphene on boron
nitride (BN) substrates\cite{Menno,Moon,Chizhova},
or by impurity doping\cite{Song}.

In recent years, there has been an extensive searching for
alternative monolayer systems similar to graphene that would be more
appropriate for nanoelectronics. One of the most recent examples is
transition metal dichalcogenides
(TMD) monolayers\cite{Frindt,Splendiani,Radisavljevic,Heine,Ataca,Gomez,Taniguchi1,Yuan,Lee},
two-dimensional monolayers with honeycomb lattice structures
similar to graphene but with gapped subband stuctures
of multi-valleys located at the K and K$^{\prime}$ of the Brillouin zone.
In contrast to graphene, a TMD monolayer can be described
effectively by the massive Dirac equation instead of the massless one.
Typically the energy band gap of TMD monolayers can be
relatively large, for instance, 1.35~eV for W${\rm S}_2$ or 1.83~eV
for Mo${\rm S}_2$. Additionally, they offer certain advantages
over conventional semiconductors, such
as large spin-orbit (SO) coupling, which makes them promising
candidates for spintronics applications.

Spintronics of conduction electrons is an
emerging field due to its applications from logic to storage devices
with high speed and very low power dissipation
\cite{Loss1998,Zutic2004,Wolf2001}. Manipulating the spin
information offers the possibility to scale down spintronic devices
to the nanoscale and is favorable for applications in quantum
computing \cite{Awschalom2002,Awschalom2007,Heedt2012}. Various SO
effects provide a promising way to spin manipulation in
two-dimensional (2D) electron gases\cite{Winkler2003,Meier2007}.
Particularly, the Rashba SO interaction is of importance in
spintronic devices, such as spin field-effect transistors
\cite{Datta1990,Bandyopadhyay2004,Koo2009}.

In the current work, we propose a way of
manipulating the quantum transport properties of spin-polarized
electrons in nanoribbons made of massive Dirac materials, by using
experimentally achievable finger-gate potential
structures \cite{Tang2012}. It is well known that a single
impurity in open 2D quantum systems at the
presence of magnetic field, may yield quasibound states (QBSs) below
the subband bottoms of the energy spectrum, which can lead to strong
backscattering of the conduction electrons, resembling evanescent scattering
effects \cite{Vargiamidis2005,Gudmundsson2005,Tang2005}.  We shall
demonstrate the role of the QBSs in Dirac nanoribbons under an
in-plane magnetic field. We show that the QBSs lead to dips in the
conductance, suppressing the spin-polarized electronic
current of the nanoribbons, similarly to localized
evanescent modes. The position of the dips can be controlled by the
polarity (sign) of the finger-gate potential, offering the
possibility for the realization of a spin-switch. The effects of the
QBSs can be detected in the low energy regime via scattering
processes involving one subband which we identify as intra-subband
scattering and in the high energy regime, via scattering processes
involving different subbands which we identify as inter-subband
scattering. These processes can involve either flipping of the
electron spin (spin-flip), or not (spin-nonflip),
allowing additional control over the spin polarized electronic
transport via partial finger-gate structures.

\section{Model}

\begin{figure}
 \includegraphics[width=\columnwidth,angle=0,scale=1.0,clip=true]{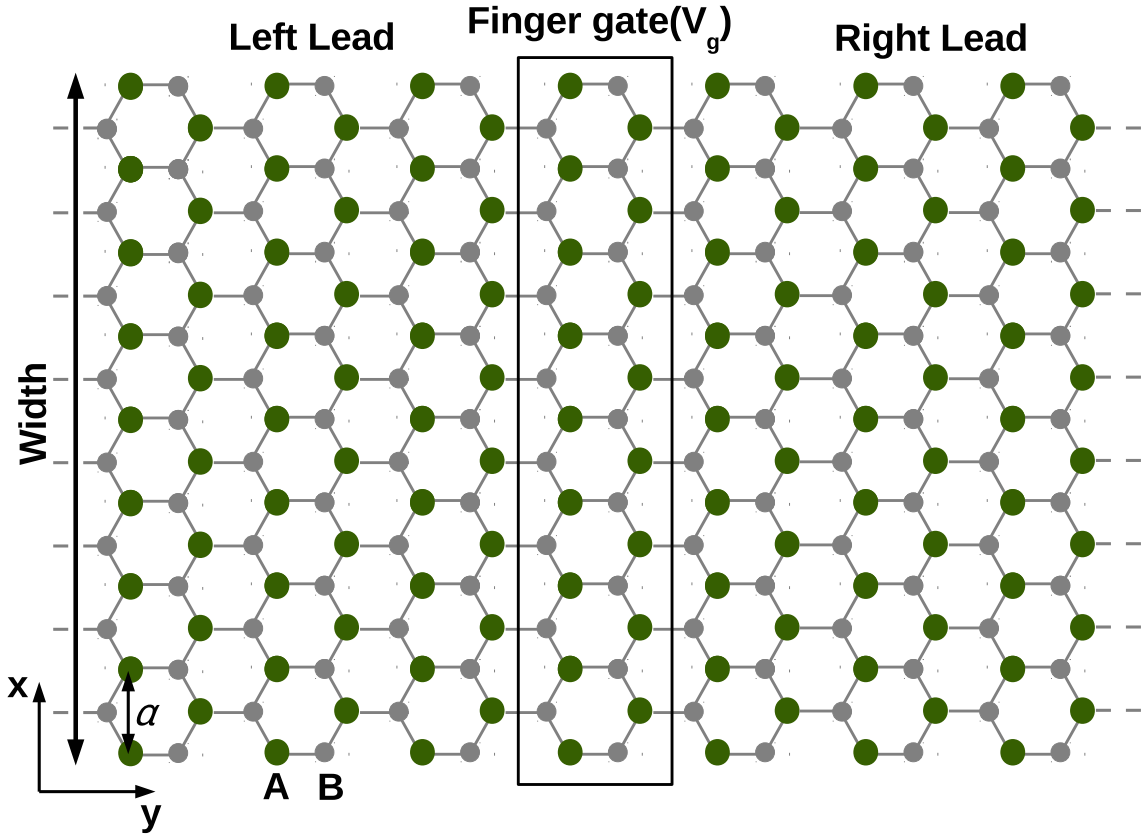}
 \caption{Schematic of the armchair nanoribbon for which we perform our calculations.
 The massive Dirac material posseses a honeycomb lattice structure
 composed of A and B sublattices with different on-site potentials +V and -V.
 We fix the width of the nanoribbon to $8a$, where $a$ is the lattice constant.
 The area inside the rectangular box indicates the unit cell where we
 impose the finger-gate potential $V_g$. The transport calculation is performed
 by treating the rest of the nanoribbon as the left and right leads (source and drain).}
  \label{fig1}
 \end{figure}

In this section, we introduce a general tight-binding model to
calculate the transport properties in nanoribbons made of Dirac
materials influenced by an external in-plane magnetic field. The
general equation that defines Dirac materials is the massive Dirac
equation\cite{dloss,Xu,Guinea,Andor,Abergel,Levitov}
\begin{equation}
\label{h_0}
{\cal H}_{0} = \hbar\upsilon_f(k_x \tau_z \sigma_1 + k_y\sigma_2) + V\sigma_3
\end{equation}
where $\upsilon_f$ is the Fermi velocity of electrons depending on
the material under investigation, and $\sigma_i$ are the Pauli
matrices. Symbol $\tau_z=\pm1$ denotes the non-equivalent valleys
that are present in graphene at the corners of its hexagonal
Brillouin zone, and other Dirac materials like TMD monolayers. \eq{h_0}
describes relativistic particles with mass $V$, where the speed of
light $c$ is replaced by $\upsilon_f$. Different values of $V$
classify different materials for example $V=0$~eV and $\upsilon_f
\approx 10^6$~m/s correspond to graphene while finite $V$ could
correspond to TMD monolayers and graphene on BN
substrates\cite{Chizhova,dloss,Guinea,Andor}.

A simple way to generate both Rashba and Zeeman
effects is via applying a rotating in-plane magnetic field with
tunable strength $B_n$ and rotation period $\lambda_n$ that can be
achieved experimentally by nanomagnets placed at the nanoribbon edges
\cite{dloss}. The resulting Rashba term is
\begin{equation}
\label{h_rn_c} {\cal H}_{Rn} = -\alpha_{Rn}s_{z}\sigma_{2}
\end{equation}
where $s_{z}$ is the Pauli
matrix while the
Rashba interaction strength is given by
$\alpha_{Rn}=\hbar\upsilon_f\pi/\lambda_n$. Alternatively this Rashba effect can be introcuced
in the system by considering an electric field along $\hat{\mb x}$.
The rotating in-plane magnetic field induces also a
Zeeman term
\begin{equation}
\label{h_z_c}
{\cal H}_{Z}=\Delta_{Z}s_{x}\,
\end{equation}
where the Zeeman interaction strength is given by $\Delta_{Z}=g
\mu_B B_{n}/2$. Notice that the spin quantization axis determined by
the Rashba term ($\hat{\mb z}$) is perpendicular to the one
determined by the Zeeman term ($\hat{\mb x}$). This
feature is important for the formation of QBSs at the edges of the
subbands and can be achieved by different configurations of
external magnetic and electric fields \cite{Tang2012}.

In order to perform our numerical calculations, we
transform \eq{h_0} to  an effective Hamiltonian
in the tight-binding formulation, consisting of a honeycomb
lattice with an onsite staggered potential described by
\begin{equation}
\label{h_0_tb} H_0= \sum_{i}\epsilon_{i}c_{i\mu}^\dagger c_{i\mu}+
\sum_{\langle i,j\rangle}t_{i,j} (c_{i\mu}^\dagger c_{j\mu}
+c_{j\mu}^\dagger c_{i\mu} )\, ,
\end{equation}
where $c_{i\mu}^\dagger$ ($c_{i\mu}$) is the creation (annihilation)
operator for spin $\mu$ at site $i$ while $t_{i,j}$ is a uniform
nearest neighbor hopping between all the lattice sites.
The onsite potential $\epsilon_i$ is $V$ and $-V$
on $A$ and $B$ sublattice sites respectively with $V=0$
corresponding to graphene. The staggered potential breaks the
inversion symmetry resulting in a gap $2V$ at the Fermi energy for
the graphene nanoribbons that were originally metallic. At low
energies the effective tight-binding Hamiltonian \eq{h_0_tb} is
equivalent to \eq{h_0} with $\hbar\upsilon_f=\frac{t\sqrt{3}a}{2}$
where $a$ is the lattice constant. In this sense \eq{h_0_tb} can be
thought as a numerical version of the massive Dirac equation. For
numerical stability we fix $t=1$~eV and $V=0.830$~eV for the numerical
calculations we present in the manuscript which results in a gap
that is comparable to the gaps observed in TMD monolayers. However our
findings concern any value of $V$ which could characterize a wide
range of Dirac materials. We provide the general conditions to
investige the QBS effects in various Dirac materials.

In order to constract a coordinate independent form of \eq{h_rn_c} in the tight-binding formulation
we modify the full Rashba interaction term originated from an
external electric field along $\hat{\mb z}$ \cite{dloss,Kane},
\begin{equation}
\label{h_r_tb} H_{R}=\frac{3i\alpha_{Rn}}{4} \sum_{\langle
i,j\rangle,\mu,\mu^{'}} \left[c_{i\mu}^\dagger \left(
\textbf{\emph{e}}_{ij} \times \textbf{\emph{e}}_z \right)
\textbf{\emph{s}}_{\mu,\mu^{'}} c_{j\mu^{'}} \right] ,
\end{equation}
where
$\textbf{\emph{s}}=({s_x,s_y,s_z})$ is the spin vector with $s_i$ being the Pauli matrices,
while the unit vector
$\textbf{\emph{e}}_{ij}$ points along the bond connecting nearest
neighbor sites i and j and $\textbf{\emph{e}}_z$ points along
$\hat{\mb z}$. By considering only the part of \eq{h_r_tb} that contains $s_{x}$
and replacing it with $s_{z}$ \cite{dloss} we can simulate ${\cal H}_{Rn}$ as follows
\begin{equation}
\label{h_rn_tb} H_{Rn}=\frac{3i\alpha_{Rn}}{4} \sum_{\langle i,j
\rangle,\mu,\mu^{'}} \left[c_{i\mu}^\dagger \left(
\textbf{\emph{e}}_{ij} \times \textbf{\emph{e}}_z \right)
s_{z,\mu,\mu^{'}} \hat{\mb x} c_{j\mu^{'}} \right] ,
\end{equation}
We fix the Rashba interaction strength to
$\alpha_{Rn}=15$~meV which corresponds to $\lambda_n=74$~nm
for all the calculations presented in the manuscript.

The Zeeman term \eq{h_z_c} is introduced in the tight-binding model via
\begin{equation}
\label{h_z_tb}
  H_{Z}=\Delta_{Z} \sum_{i,\mu,\mu^{'}} c_{i\mu}^\dagger s_{x,\mu,\mu^{'}} c_{i\mu^{'}} ,
\end{equation}
where we fix also $\Delta_{Z}=$0.05~meV, corresponding to
$B_n=0.86$~T. The total effective tight-binding
Hamiltonian of the system under investigation is
\begin{equation}
\label{h_tb}
  H = H_0+H_{Rn}+H_{Z} \, .
\end{equation}
We consider hard-wall boundary conditions along
$\hat{\mb x}$ forming nanoribbons with armchair edges in order to
avoid edge effects. We fix the nanoribbon width to 8$a$
corresponding to eight hexagons along $\hat{\mb x}$ perpendicular to
the transport direction $\hat{\mb y}$ (\fig{fig1}).

The energy dependence of the conductance $G$, in units of
conductance quantum $e^2/h$, is calculated via a recursive Green's
function method  within the framework of Landauer-B\"{u}ttiker
formalism \cite{li,li1}. To analyze the evanescent effects
associated with the quasibound-state feature, we utilize a
finger-gate potential as a scatterer, simulated by a uniform onsite
potential $V_g$ on every site inside one unit cell of the armchair
Dirac nanoribbon in the effective tight-binding model, as shown in
\fig{fig1}. We also consider partial finger-gates by placing $V_g$
only on a few sites inside the unit cell. The
corresponding partial density of states (PDOS) can be expressed in
terms of the Green function element on site $i$ inside the scatterer
${\cal G}\left(i,i,E\right)=(E-H_S-\Sigma_L-\Sigma_R)^{-1}$ with $E$
being the incident energy, $H_S$ being the scatterer Hamiltonian,
and $\Sigma_L$$(\Sigma_R)$ being the self energy of the left(right)
lead, as
\begin{equation}
\label{pdos}
  \mathrm{PDOS} = -\frac{1}{\pi} \Im\left[ {\cal G}\left(i,i,E\right)\right]\,
  .
\end{equation}
This quantity is useful for illustrating the QBS
feature when the site $i$ is considered as a substitutional
scatterer.

\section{Subband Structure}

\begin{figure}
\includegraphics[width=\columnwidth,angle=0,scale=1.0,clip=true]{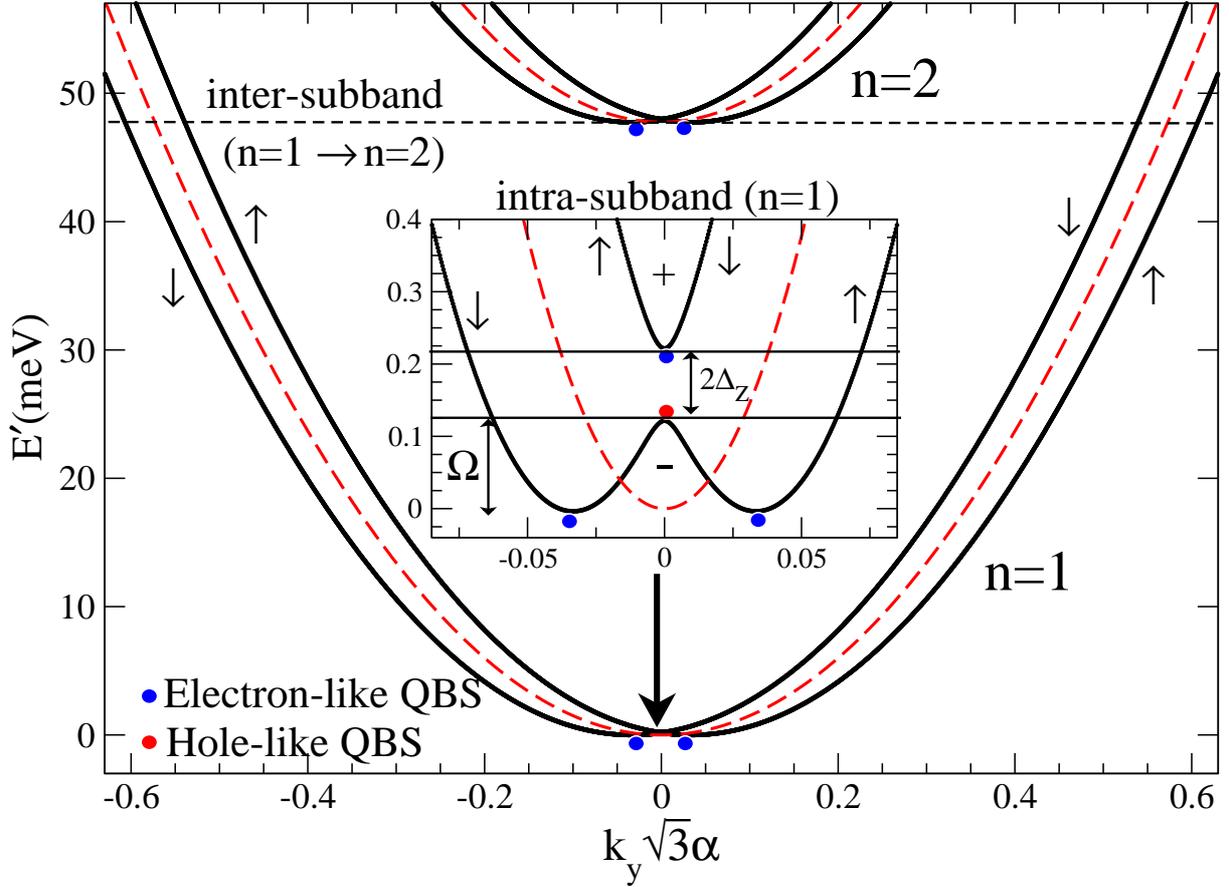}
\caption{Subband energy structure $E^{\prime}$
versus $k_y\sqrt{3}a$ for a Dirac armchair nanoribbon of width
$W=8a$ (red dashed curves). The black curves are the subband structure with
Rashba strength $\alpha_{Rn}=15$~meV and Zeeman strength
$\Delta_{Z}=$0.05~meV. The Rashba term lifts the spin degeneracy
along the transport direction $\hat{\mb y}$ creating two branches
with opposite spin polarizations (opposite arrows) for both $n$=1
and $n$=2 subbands. The Zeeman term opens a SO gap $2\Delta_Z$
between the branches belonging to the same subband, at $k_y = 0$ as
shown in the inset. The red (blue) dots at the edges of the branches
denote the HQBS and EQBS that are known to lead to evanescent scattering
effects. Two types of scattering through the QBSs are possible,
intra-subband occurring inside the SO-Zeeman gap (inset) and
inter-subband from $n$=1 to $n$=2 subbband branches creating
additional scattering possibilities. These processes can be either
of spin-flip or spin-nonflip type.}
 \label{fig2}
\end{figure}

In this section, we shall analyze the spectrum of the Dirac
nanoribbon. Moreover, we shall show analytical expressions to
describe the lowest conduction subbands in the presence of the
Rashba spin-orbit interaction and the Zeeman effect due to the
presence of the in-plane magnetic field.

In the absence of magnetic field, when only \eq{h_0} is
considered, the lowest conduction subbands, which we characterize by
$n$, follow a parabolic form for small $k_y$ corresponding to the massive Dirac
fermions
\begin{equation}
\label{disp}
  E(k_y)= \varepsilon_n + \frac{(\hbar\upsilon_f k_y)^2}{2\varepsilon_n}
\end{equation}
where $\varepsilon_n$ is the minimum energy of each subband.

The numerical subband structure can be seen in \fig{fig2}, where we
plot the shifted incident energy
\begin{equation}
E^{\prime} = E - \varepsilon_1
\end{equation}
as a function of rescaled longitudinal wave number $k_y\sqrt{3}a$.
We use the energy $E^{\prime}$ for all the
numerical calculations in the manuscript.

The red dashed curves in
the main figure are the two lowest conduction subbands of $H_0$
denoted by $n=1$ and $n=2$, with minimum energies
$\varepsilon_1=V=830$~meV, $\varepsilon_2=877.8$~meV in \eq{disp}.
In the absense of magnetic field every subband is doubly degenerate in spin (Kramers degeneracy), corresponding to right and left propagating
modes.

The inclusion of the Rashba term
\eq{h_rn_c}, described in tight-binding
terms by \eq{h_rn_tb}, causes a spin-splitting along the transport direction $k_y$ creating two
different branches for every subband $n$ represented by the black
curves in \fig{fig2}. The branches are characterized by opposite
spin polarizations, denoted by $\uparrow$ and $\downarrow$ in \fig{fig2}. However the
Kramers degeneracy of the spin is preserved in this case. The Rashba
spin-splitting mechanism is equivalent to a spin dependent shift of
the wave vector $k_y \rightarrow k_y\pm
\frac{\alpha_{Rn}}{\hbar\upsilon_f}$ resulting from the fact that the
spin $s_{z}$ is a good quantum number for the Hamiltonian \eq{h_0}\cite{dloss}.

The Kramers degeneracy of the spin is lifted by
the Zeeman term \eq{h_z_c}, since the corresponding spin quantization axis ($\hat{\mb x}$)
is perpendicular to the one determined by
the Rashba term \eq{h_rn_c} ($\hat{\mb z}$). Consequently a SO gap
$2\Delta_Z$=0.1~meV opens at $k_y=0$ between the opposite spin branches
originally created by the Rashba interaction. The SO gap can be
seen in the inset of \fig{fig2}. Inside the gap there is one
right and one left propagating mode characterized by opposite spin
polarizations meaning that electrons traveling in opposite
directions have opposite spins, creating a helical regime. This
feature is important in quantum transport measurements since it
leads to spin polarized current. The actual spin orientation along
the branches for a given $k_y$ depends on the energy difference
between the opposite spin branches, however the electrons are almost
fully spin polarized along the $\hat{\mb z}$ direction (parallel or
antiparallel) as long as the incident energy of the electrons coming
from the leads is tuned higher than the middle of the
SO gap \cite{dloss}.

The subband structure in the inset of \fig{fig2}
resembles the energy dispersion of an electron in a homogeneous 2D
quantum wire at the presence of an in-plane magnetic
field \cite{Tang2012}. By plugging the Rashba-shifted wave vector
$k_y \rightarrow k_y\pm \frac{\alpha_{Rn}}{\hbar\upsilon_f}$, and
the Zeeman effect (gap) in \eq{disp} we obtain an analytical formula
that describes the branches of the subband $n$, given by
\begin{eqnarray}
 \label{E_snky}
E^{\sigma}_{n}(k_y) &=& \varepsilon_n^{\prime} +
\frac{(\hbar\upsilon_f k_y)^2}{2\varepsilon_n}
  +\sigma \sqrt{ \left(\frac{\alpha_{Rn}^{\prime}\hbar\upsilon_fk_y}{\varepsilon_n} \right)^2 + (\Delta_Z)^2 }
\end{eqnarray}
where $\alpha_{Rn}^{\prime} = \eta \alpha_{Rn}$ with the factor
$\eta = 9/8$ obtained by fitting the numerical subband structure in
\fig{fig2}. This fitting factor is related to the honeycomb lattice
morphology. The symbol $\sigma=\pm$ indicates the two different
branches with opposite spin polarizations. The energy
$\varepsilon_n^{\prime}$ is the corresponding modified ground state
electronic energy in the quantum wire model
\begin{equation}
\label{E_n}
\varepsilon_n^{\prime} = \varepsilon_n +
\frac{(\alpha_{Rn}^{\prime})^2}{2\varepsilon_n}\, .
\end{equation}
The corresponding effective Rashba
interaction strength is $\alpha_{Rn}^{\prime}/\sqrt{2\varepsilon_n}$ while the
Zeeman strength remains the same $\Delta_Z$. The corresponding kinetic energy of the electron would be
$K_n = (\hbar\upsilon_f k_y)^2/(2\varepsilon_n)$.
The simple quantum wire model with in-plane magnetic field, with the effective parameters mentioned above
could effectively describe the electron behavior in Dirac
nanoribbons at the presence of a rotating in-plane magnetic field.

Based on \eq{E_snky}, we define the spin-resolved branches of the subbands as
$\arrowvert n,\sigma \rangle$, so that $\arrowvert 1,- \rangle$ and
$\arrowvert 1,+ \rangle$ ($\arrowvert 2,- \rangle$ and $\arrowvert 2,+ \rangle$) represent, the lower and
upper branches of the first (second) subband $n$=1 ($n$=2).

The red and blue dots in \fig{fig2} at the edges of the subband branches, correspond, to the
hole-like QBS (HQBS) and electron-like QBS (EQBS), respectively, formed
under the influence of the magnetic field. Under certain conditions the QBSs are known to induce backscattering in mesoscopic
conductors, similarly to localized evanescent modes \cite{Vargiamidis2005,Gudmundsson2005,Tang2005}.
A finger-gate potential provides a convenient way to investigate the QBS effects
in Dirac materials.
The two EQBSs (blue dots) at the minima of the
branch $\arrowvert 1,- \rangle$ in the inset of \fig{fig2} are
located slightly lower than  $E^{\prime}=0$~meV, due the Zeeman
interaction. The HQBS (red dot) at the maxima
of $\arrowvert 1,- \rangle$ is located at energy $E_{\rm
HQBS}^{\prime}(\arrowvert 1,- \rangle)\simeq 0.122$~meV. In analogy
an EQBS is located at the minima of the upper branch $\arrowvert
1,+ \rangle$ at $E_{\rm EQBS}^{\prime}(\arrowvert 1,+
\rangle)\simeq0.222$~meV. The Zeeman gap is $E_{\rm
EQBS}^{\prime}(\arrowvert 1,+ \rangle)- E_{\rm
HQBS}^{\prime}(\arrowvert 1,- \rangle) = 2\Delta_Z = 0.1~meV$.

From the subband structure in \fig{fig2}, two types of scattering
processes associated with the QBSs can be distinguished. One is
intra-subband occurring between different branches belonging to same
subband $n$ for example from branch $\arrowvert 1,- \rangle$ to either the
maximum of the same branch or to the bottom of $\arrowvert 1,+
\rangle$, as shown by solid lines in the inset of \fig{fig2}.
Additionally, inter-subband scattering can occur at higher energy
between branches belonging to different subbands, denoted by a
dashed line from $n$=1 to $n$=2 in \fig{fig2}.
These transitions correspond to the scattering from either $\arrowvert 1,- \rangle$ or
$\arrowvert 1,+ \rangle$, to the subband bottoms of $\arrowvert 2,-
\rangle$. When a scattering process involves
transitions between branches characterized by opposite $\sigma$ then
it corresponds to spin-flip backscattering through the QBS. For instance the
intra-subband scattering process, $\arrowvert 1,- \rangle$ to $\arrowvert 1,+
\rangle$ is spin-flip, while the inter-subband process $\arrowvert 1,+ \rangle$ to $\arrowvert
2,+ \rangle$ in spin-nonflip. We expect spin-flip scattering
processes to occur at higher energies than the spin-nonflip ones,
offering additional possibilities to control the the
quantum transport properties of the spin-polarized electrons
through finger-gate structures.

The width of the low energy regime of the
$\arrowvert 1,- \rangle$  branch ($E_{\rm HQBS}^{\prime}(\arrowvert
1,- \rangle)-E_{\rm EQBS}^{\prime}(\arrowvert 1,- \rangle)$) denoted
by $\Omega$ in the inset of \fig{fig2} plays an important
role. Considering only the Rashba interaction, it is proportional to the
second term of \eq{E_n} namely
\begin{equation}
\label{omega}
\Omega\sim\frac{\alpha_{Rn}^2}{V},
\end{equation}
obtained by putting $\Delta_Z=0$ and $k_y=0$ in \eq{E_snky}.
Increasing $\alpha_{Rn}$ results in a quadratic
increase of $\Omega$. On the other hand, increasing $V$ results in
less steep slope of the subbands which reduces
the effect of the Rashba spin splitting along $k_y$,
effectively decreasing $\Omega$.
This behavior is retained for finite $\Delta_Z$
which simply opens a SO gap between the opposite
spin branches created by the Rashba interaction.
The interplay between $\alpha_{Rn}$
and $V$ and it's effect on $\Omega$ is
important for identifying the QBS scattering mechanism as we shall
demonstrate below.

\section{Intra-subband Scattering}

\begin{figure}
\includegraphics[width=\columnwidth,angle=0,scale=1.0,clip=true]{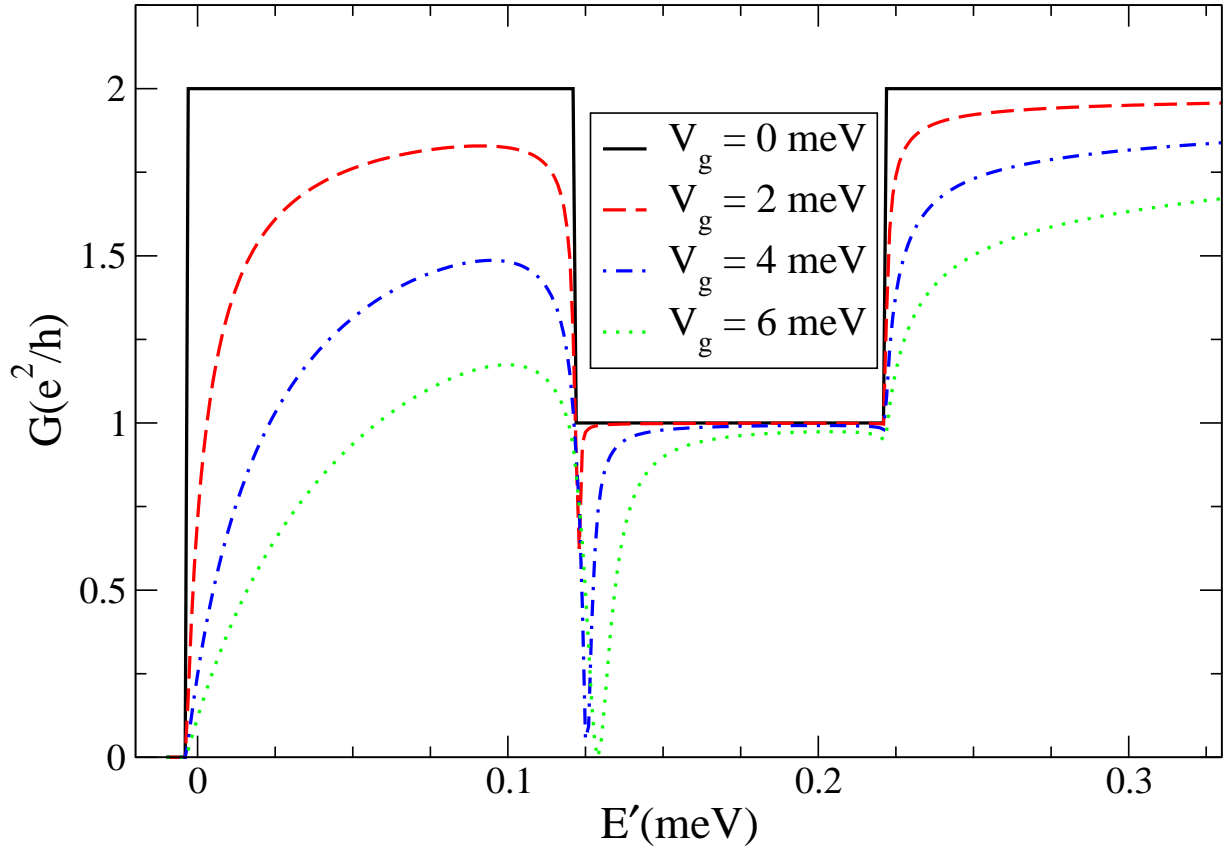}
  \caption{Conductance $G$ versus the shifted energy $E^{\prime}$ for the Dirac armchair
  nanoribbon for the two opposite spin branches of the subband $n$=1, with positive (repulsive) finger-gate potential
  of strength $V_g$. The ballistic case $V_g$=0~meV agrees with the subband structure in the inset of \fig{fig2}.
  The scattering via the hole-like QBS (HQBS) creates a sharp dip at the low-energy end
  of the SO gap. The extent and the position of the dip varies slightly with $V_g$.
  For large $V_g$ the conductance is completely suppressed ($G=0$). }
  \label{fig3}
 \end{figure}

\begin{figure}
\includegraphics[width=\columnwidth,angle=0,scale=1.0,clip=true]{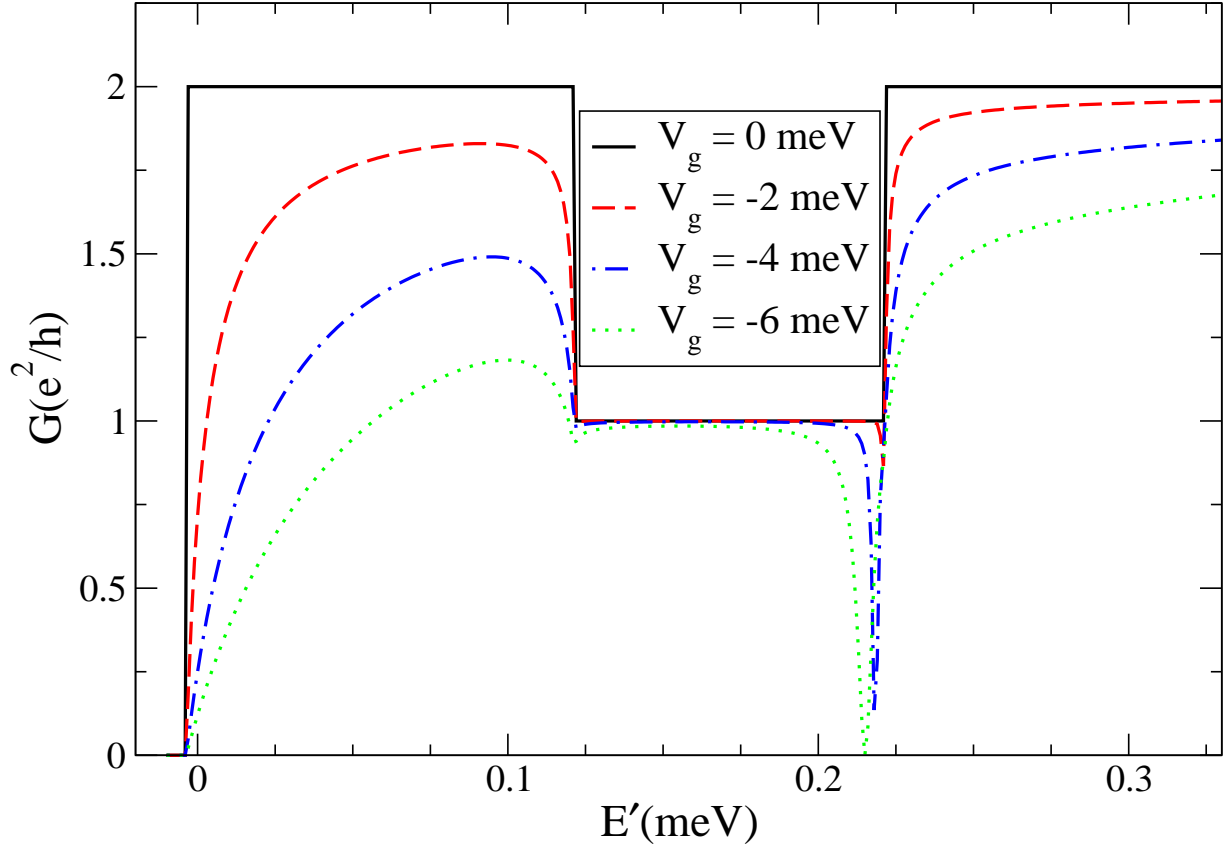}
\caption {Conductance $G$ versus
 $E^{\prime}$ for negative (attractive) $V_g$. The effect of the scattering via the
EQBS can be distinguished at the high-energy end of the SO gap where
the conductance is supressed.}
  \label{fig4}
 \end{figure}

In this section, we demonstrate the intra-subband scattering effects
within the Zeeman induced SO gap, shown in the inset of \fig{fig2}. The conductance
versus the shifted energy $E^{\prime}$ for the subband $n$=1 can be seen in \fig{fig3}
and \fig{fig4} for positive and negative finger-gate potential strengths. Also, the case
of ballistic transport without the finger-gate potential
(V$_g$=0~meV) is also shown agreeing with the
band structure in the inset of \fig{fig2} leading to maximum
conductance $G$=1 inside the Zeeman gap and to $G$=2 outside. For the case of positive potential in  \fig{fig3},
the effect of the corresponding HQBS at the low end of the SO gap near $E_{\rm HQBS}^{\prime}(\arrowvert
1,- \rangle)$,can be clearly
seen as a sharp dip suppressing the conductance.
The dip widens as
the potential strength increases while its position varies slightly
also. For large potential strengths e.g for $V_g=6$~meV the
conductance is completely suppressed ($G=0$). Inverting the finger-gate potential,
which corresponds to negative $V_g$ in \fig{fig4} a
similar dip is created at the high end of the SO gap near $E_{\rm
EQBS}^{\prime}(\arrowvert 1,+ \rangle)$ where an EQBS lies.
Positive (negative) finger-gate potential leads to evanescent
scattering via HQBS(EQBS) located near $E_{\rm
HQBS}^{\prime}(\arrowvert 1,- \rangle)$ ($E_{\rm
EQBS}^{\prime}(\arrowvert 1,+ \rangle)$) in the inset of  \fig{fig2},
suppressing the conductance near the edges of the Zeeman induced SO
gap. We notice that the intra-subband scattering involves
both spin-flip and spin-nonflip scattering processes.
The dip induced by the negative finger-gate potential is a spin-flip process since
it is caused by a transition between opposite spin branches, from
$\arrowvert 1,- \rangle$ to $\arrowvert 1,+ \rangle$.
On the other hand the dip induced by the positive finger-gate potential is a spin-nonflip process since it involves only the lower branch
$\arrowvert 1,- \rangle$.
The energy for the spin-flip process is higher
that the spin-nonflip one.

The intra-subband mechanism can be used as a spin-switch. By
fixing the Fermi energy near either $E_{\rm
HQBS}^{\prime}(\arrowvert 1,- \rangle)$ or $E_{\rm
EQBS}^{\prime}(\arrowvert 1,+ \rangle)$ at the edges of the SO gap,
and inverting the polarity of the finger-gate potential it is
possible to switch on and off the spin polarized electronic current.
This mechanism would be more appropriate for energies near $E_{\rm
EQBS}^{\prime}(\arrowvert 1,+ \rangle)$ where the electrons have
accumulated full spin polarization along the $\hat{\mb z}$
axis\cite{dloss}.

The intra-subband mechanism is not affected by the width of the
ribbon since the parabolic form (\eq{disp}) of the $n$=1 subband and
the spin splitting caused by the Rashba and Zeeman terms described
analytically by \eq{E_snky} are maintained for any ribbon width.
Moreover the intra-subband scattering can be detected also for a few
subsequent subbands like the $n$=2 in \fig{fig2}. We have derived
results similar to the ones shown in \fig{fig3} and \fig{fig4} for
the two branches of the $n$=2 subband, which can be seen in \fig{fig5}.
Also, enlarging the range of the finger-gate will
introduce multiple scattering effects, and hence reduce slightly the
conductance, but the essential transport properties will not
change.

The evanescent effects induced by the QBSs can
be observed for different values of the staggered potential $V$
which corresponds to different Dirac materials. For instance, in the $V=0$~meV limit,
which corresponds to armchair graphene nanoribbons the QBS effects
can be detected for the semiconducting
ribbons with gap, whose low energy subband structure resembles the inset of \fig{fig2} \cite{dloss1}.
For finite $V$, the extent of the dips and the level of suppression of the conductance
depends on fine-tuning the gate-potential strength $V_g$ along with
the rotation period $\lambda_n$ of the magnetic field which
determines the Rashba strength $\alpha_{Rn}$. The width $\Omega$
of the low energy regime of the $\arrowvert 1,-
\rangle$ branch (see \eq{omega}), shown in the inset of figure \fig{fig2}, plays an
important role. We have found that in order to distinguish the QBS
dips clearly $V_g$ should be much larger that $\Omega$, which can be
roughly expressed as
\begin{equation}
\label{cond} V_g\gg\frac{\alpha_{Rn}^2}{V}
\end{equation}
in terms of $\alpha_{Rn}$ and V. In
principle, for any Dirac material, a fine-tuning of the magnetic
field parameters and $V_g$ according to \eq{cond} should achieve the
desirable conductance suppression which is required for better
control of the spin-switch. Another necessary condition is that the Zeeman gap ($2\Delta_Z$) determined by the magnetic field strength $B_n$ should be sufficiently large so that the EQBS and HQBS are clearly separated. In general,
since \eq{cond} is inversely proportional to V, Dirac materials with
large gaps like TMD monolayers require smaller $V_g$ than materials with
smaller gaps, like graphene on BN substrates.

\section{Inter-subband scattering}

\begin{figure}
\includegraphics[width=\columnwidth,angle=0,scale=1.0,clip=true]{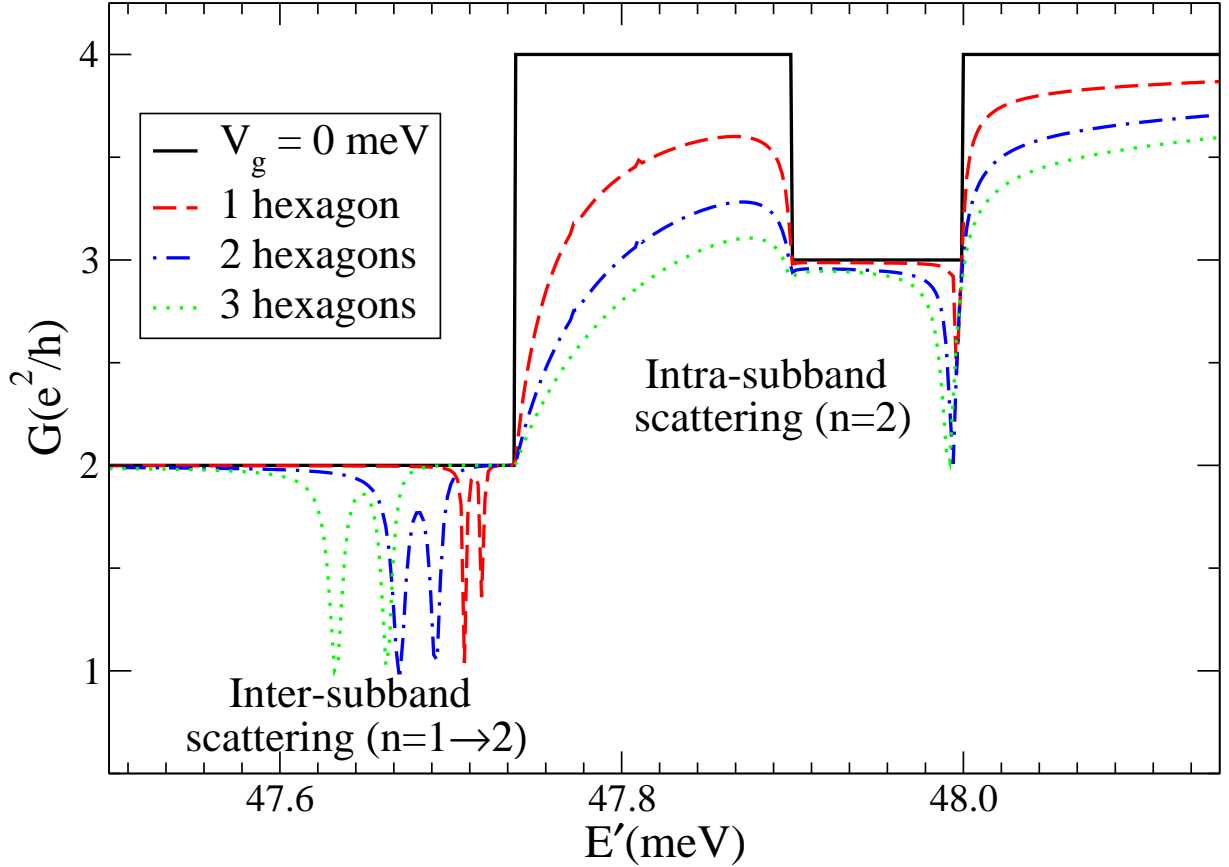}
\caption{Conductance $G$ versus $E^{\prime}$ near the subband $n=2$
for a partial finger-gate potential covering one, two, and three hexagons, with
negative potential strength $V_g=-20$~meV. The intra-subband scattering is retained
at the high energy end of the Zeeman induced SO gap between the $n=2$ branches, leading to a
sharp dip like in the case of the $n=1$ subband branches. The inter-subband scattering mechanism can be seen
before the conductance jumps from $G=2$ to $G=4$, as double dip structures, coming from the scattering
either from the $\arrowvert 1,+ \rangle$  or $\arrowvert 1,- \rangle$
branches, via the two EQBS at the bottoms of the $\arrowvert 2,-
\rangle$ branch. These transitions are spin-flip and spin-nonflip respectively.}
  \label{fig5}
 \end{figure}

An additional scattering mechanism is possible between branches corresponding to
different subbands $n$, denoted by a dashed line between $n$=1 and $n$=2 in the
subband structure shown in \fig{fig2}. Two scattering processes are possible, from either $\arrowvert 1,- \rangle$ or
$\arrowvert 1,+ \rangle$, to the subband bottoms of $\arrowvert 2,-
\rangle$. These processes create
additional possibilities to control the electronic transport via
partial finger-gate structures.

Since the inter-subband effects manifest due to electron-like QBS(EQBS)
we need to use negative finger-gate potential in order to detect them.
Additionally in order to enhance the inter-subband mechanism, one needs to break the symmetry of the ribbon along the $\hat{\mb x}$ direction,
perpendicular to the transport\cite{li1}. To this end, we
utilize a partial finger-gate potential constructed by choosing only
a part of the complete finger gate consisting of one, two, or three
hexagons, to place the onsite potential $V_g$. The corresponding
partial finger-gate widths would be 1$a$, 2$a$, or 3$a$,
respectively.  The conductance as a function of the shifted energy
$E^{\prime}$ is shown in \fig{fig5} for different partial
finger-gate potential areas for constant gate-potential strength
$V_g=-20$~meV.  The effect of the intra-subband scattering is
retained for the $n=2$ branches leading to a sharp dip at the high
end of the SO-Zeeman gap as in the case of the scattering occurring
between the branches of the subband $n=1$.

In \fig{fig5}, the effects of the inter-subband scattering can be
seen as a pair of sharp dips before the conductance jumps from
G=2 to G=4 at $E^{\prime}=47.7~meV$ due to the two
additional channels opening from the two branches $\arrowvert 2,-
\rangle$ and $\arrowvert 2,+ \rangle$.  The two dips in each pair
are caused by different types of scattering, namely the spin-flip
and spin-nonflip scattering mechanisms. The dip occuring at
higher energy comes from the spin-flip transition between
the $\arrowvert 1,+ \rangle$ branch and the two EQBS lying at the
bottoms of the $\arrowvert 2,- \rangle$ branch (blue dots) in \fig{fig2}.
On the other hand, the dip with the lower energy in each pair comes
from the spin-nonflip transition between the branches $\arrowvert
1,- \rangle$ and $\arrowvert 2,- \rangle$. The energy separation
between the dip pair is getting larger by increasing the area of the
partial finger-gate potential.

We notice that when the finger-gate occupies two
and three hexagons, the conductance at each dip is suppressed
approximately at $G=1$ leaving practically only one open channel for the
conduction of electrons coming either from the $\arrowvert 1,-
\rangle$ or the $\arrowvert 1,+ \rangle$ branch that are
characterized by opposite spin polarizations. This fact implies that
the remaining current at each dip is spin polarized, with
opposite spin polarization between the dips of each pair.
The inter-subband induced dips cannot be detected with positive
finger-gate potential.

\begin{figure}
\includegraphics[width=\columnwidth,angle=0,scale=1.0,clip=true]{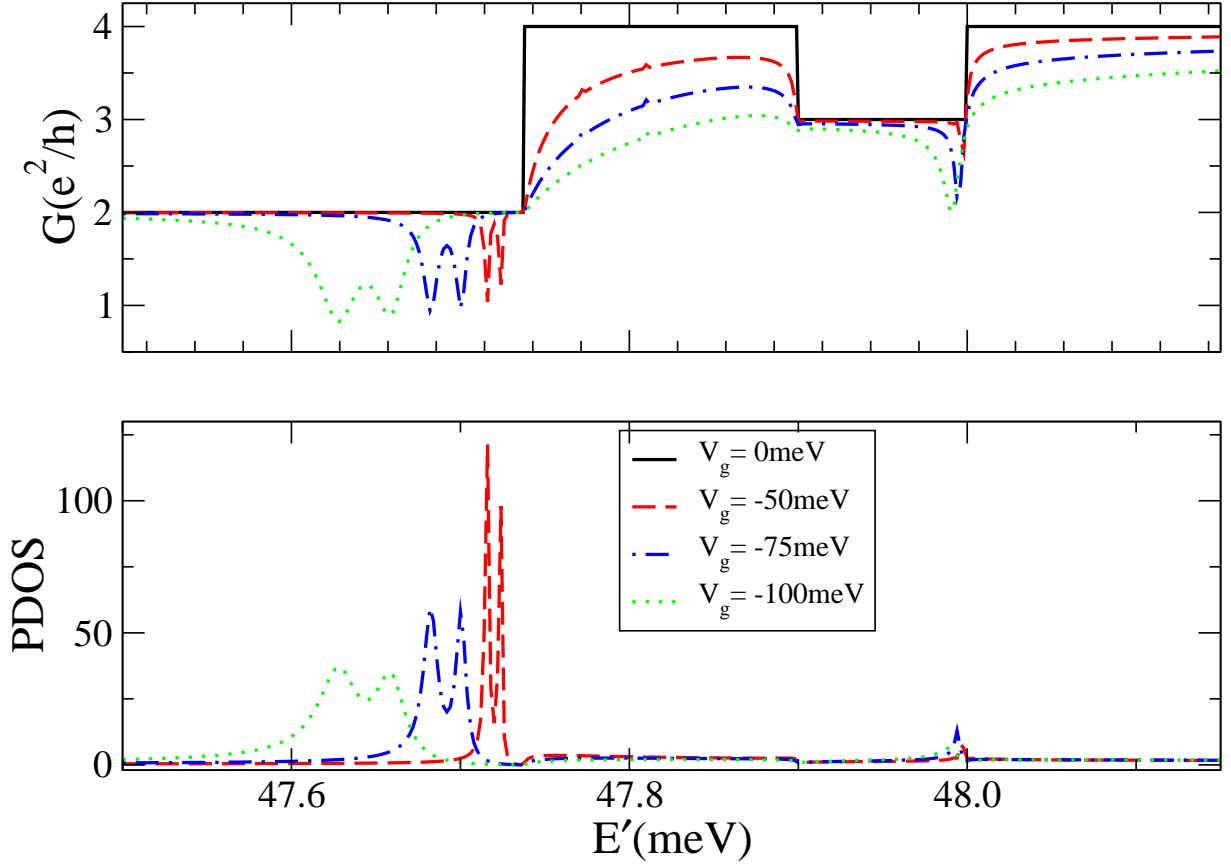}
\caption{Conductance $G$ versus $E^{\prime}$ for
one impurity at the edge of the ribbon for different potential
strengths $V_g$. The effects are very similar to using a partial
finger gate where both the inter and intra-subband scattering
processes can be distinguished. Increasing $V_g$ results in similar effect
to using larger area of the partial finger-gate.
In the bottom figure the PDOS at the position of the scattering potential
becomes clearly localized at the energies where the conductance dips occur.}
  \label{fig6}
 \end{figure}

Due to these features the inter-subband scattering
could be utilized to create and control spin polarized
electronic transport. For instance, by
inverting the polarity of finger-gate from positive to negative for
a constant energy at the position where the dips occur, creates spin polarized
transport with $G=1$ for sufficiently large finger-gate potential widths. In addition,
by varying the Fermi energy of the Dirac nanoribbon between the dips in each pair the spin
polarization of the remaining current is inverted. An alternative
method to generate similar effects is to consider one single impurity at
one-site at the edge of the nanoribbon. The result can be seen in
\fig{fig6} for various potential strengths. The PDOS shown in the bottom
becomes maximum at the position of the conductance dips,
indicating the quasi-localized QBSs. Using larger potential is roughly equivalent
to using larger area of the partial finger-gate, both resulting in stronger
backscattering. We note that the inter-subband  mechanism
is valid for any nanoribbon width since it involves transitions between
different subbands which become more dense for wider nanoribbons.
The level of control of the spin polarized transport for different $V$ corresponding to
different Dirac materials depends on the fine-tuning of the magnetic
field in conjunction with the finger-gate potential strength.

\section{Concluding Remarks}

We have proposed an experimentally achievable finger-gate
manipulation of the spin-polarized electronic transport in
nanoribbons made of Dirac materials under an in-plane magnetic field.
A numerical tight-binding model has been utilized to calculate
the subband structures and spin-resolved electronic
transport involving intra-subband and inter-subband transitions.

The QBSs at the edges of the subband branches due to the Rashba and Zeeman
effects result in conductance dips at the ends of the induced SO
gap, via intra-subband transitions. The dip positions alternate
between the low and the high end of the SO gap depending on the
polarity of the finger-gate, allowing the realization of a
spin-switch that can control the spin-polarized electronic
transport. Additionally, inter-subband transitions are possible
between different subbands. These transitions can be either spin-flip or
spin-nonflip, creating pairs of dips in the conductance, since the
energy required by each type of process is different.
For any Dirac material, a combined fine-tuning of
the magnetic field in conjuction with the finger-gate potential strength $V_g$
could provide the desirable optimal level of
control for the spin-switch.

To summarize we have demonstrated the control of transmission
spectra through Dirac nanoribbons using a tunable finger-gate setup
to dramatically influence the quantum transport. Conductance drops
can be created around the upper or lower bound of the Zeeman induced
SO gap depending on the polarity of the finger-gate potential. This
feature opens up a possibility for building quantum switch devices
based on Dirac materials. We provide also additional ways to
control the transmission spectra in the high energy regime via
partial finger-gate structures.

\section{Acknowledgements}
We are grateful to Victor A. Gopar for valuable discussions. This
work was supported by Ministry of Science and Technology, Taiwan
through No.\ MOST 103-2112-M-239-001-MY3, and National Science
Council through No.\ NSC 102-2112-M-009-009-MY2.

\end{document}